\documentclass[12pt]{article}
\usepackage{epsfig}

\newcommand{\be}{\begin{equation}}
\newcommand{\ee}{\end{equation}}
\def\lsim{\:\raisebox{-0.5ex}{$\stackrel{\textstyle<}{\sim}$}\:}
\def\gsim{\:\raisebox{-0.5ex}{$\stackrel{\textstyle>}{\sim}$}\:}

\begin{document}  
\renewcommand{\thefootnote}{\fnsymbol{footnote}}

\pagestyle{plain}
\begin{flushright}
TUM-HEP-378-00 \\
June 2000
\end{flushright}

\vspace{1cm}

\begin{center}

{\large \bf New Constraints on ``Cool'' Dark Matter}

\vspace{7mm}
Manuel Drees and David Wright \\

\vspace*{5mm}

{\it Physik Department, TU M\"unchen, James Franck Str., D--85748
Garching, Germany}
\end{center}

\vspace{2cm}

\begin{abstract}

\noindent 
It has been suggested that a sterile neutrino $\nu_s$ which mixes with
standard neutrinos can form nonthermal ``cool'' Dark Matter if its
mass and mixing angle fall in the ranges 0.1{\thinspace}keV~$\lsim m_s
\lsim$~10{\thinspace}keV and $10^{-10} \lsim \sin^2 \theta \lsim
10^{-4}$, respectively. We point out that the required mixing makes
these heavy neutrinos unstable. The dominant decay mode is into three
light neutrinos, but the most stringent constraint comes from the
non-observation of radiative decays into a single light neutrino and a
photon. Moreover, we point out that the density of thermal relics of
such $\nu_{s}$s would be too high, unless the reheating temperature
after inflation was below $\sim$10{\thinspace}GeV.

\end{abstract}

\clearpage

Recently Shi and Fuller suggested \cite{1} a new particle physics
candidate for the Dark Matter in the Universe.  They note that, if a
significant lepton-antilepton asymmetry $\Delta L$ existed in the
early Universe and was reflected in a net neutrino number, and if
there is a sterile neutrino species $\nu_{s}$ with an appropriate mass
and mixing angles, the standard neutrino excess is converted into a
sterile neutrino excess via resonant interactions with the thermal
plasma at a temperature of $\sim$10{\thinspace}MeV.  Since neutrinos
with the lowest energy will be converted first, until $\Delta L$ is
essentially used up, the resulting sterile neutrinos, while mostly
relativistic, will be significantly ``colder'' than the standard
neutrinos. More exactly, their energy distribution will be nonthermal,
with average energy a fraction of that of the parent (standard)
neutrinos. This nonthermal ensemble of sterile neutrinos $\nu_{s}$
would survive to this day, and constitutes a suitable Dark Matter
candidate for mass and mixing angle parameters in the ranges
0.1{\thinspace}keV~$\lsim m_s \lsim$~10{\thinspace}keV and $10^{-10}
\lsim \sin^2 \theta \lsim 10^{-4}$, respectively.  Such ``cool'' Dark
Matter might be compatible with studies of structure formation, while
hot Dark Matter seems to be excluded.  An explicit realization of this
idea, where $\nu_s$ is a supersymmetric axino, has been suggested by
Chun and Kim \cite{2}.

The authors of refs.~\cite{1,2} seem to have overlooked the fact that
the mixing of $\nu_s$ with standard neutrinos will cause the $\nu_s$
to decay, and will induce interactions between $\nu_s$ and the thermal
plasma. Obviously, $\nu_s$ can only form Dark Matter if it is
sufficiently long-lived. In addition, it will only form {\em cool}
Dark Matter if the density of thermal $\nu_s$ relics is sufficiently
small. We find that the first condition imposes a significant
constraint on parameter space that is independent of the thermal
history of the Universe. The second condition excludes the model
altogether, if the post-inflationary Universe was ever hot enough for
$\nu_s$ to have been in thermal equilibrium. Even for the smallest
mixing angle $\sin^{2}\theta \sim 10^{-10}$ compatible with efficient
conversion of a lepton asymmetry $\Delta L$ into $\nu_s$s, the $\nu_s$
freeze-out temperature only amounts to
$\sim${\thinspace}10{\thinspace}GeV.  ``Cool'' Dark Matter would thus
require the reheating temperature after inflation to be below this
value.

We start our discussion with an analysis of $\nu_s$ decays. Since we
are interested in sterile neutrinos with mass $m_s \lsim 10${\thinspace}keV,
the only possible tree-level decay is the one into three light Standard
Model (SM) neutrinos, $\nu_{s} \rightarrow 3\nu$, which
proceeds through the exchange of a $Z$-boson. The $Z \nu_s \bar{\nu}$
coupling is simply $\sin \theta$ times the $Z \nu \bar{\nu}$ coupling
of the SM, where $\theta$ is the $\nu - \nu_s$ mixing angle.
The corresponding partial width is given by
\be \label{e1}
\Gamma(\nu_s \rightarrow 3 \nu) = \frac {G_F^2 m_s^5} {192 \pi^3}
\sin^2 \theta,
\ee
where $G_F$ is the Fermi constant, and we have summed over all three
generations of SM neutrinos. $\sin^{2}\theta =
\sum_{i} \sin^{2} \theta_{i}$, where $\theta_i$ is the
mixing angle between $\nu_s$ and $\nu_i$, is the effective
mixing angle to standard model neutrinos.

An obvious constraint on any particle physics candidate for Dark
Matter is that its lifetime should exceed the age of the Universe,
$\tau(\nu_{s}) \gsim 5 \cdot 10^{17}$ sec. However, in the case at hand a
stronger constraint can be derived, since the mixing between $\nu_s$
and the light neutrinos also induces radiative decays at the one--loop
level, with branching ratio \cite{3}
\be \label{e2}
B(\nu_s \rightarrow \nu \gamma) = \frac {27 \alpha_{\rm em}} {8 \pi}
\simeq 8 \cdot 10^{-3},
\ee
where $\alpha_{\rm em}$ is the fine structure constant. These decays
would add a monochromatic line at energy $E = m_s/2$ to the diffuse
background of hard UV or soft X-ray photons. There are stringent
observational bounds on such anomalies \cite{4}, leading
very conservatively to
the requirement\footnote{This constraint is weaker than the
corresponding one in ref.~\cite{4} by a factor of the branching ratio
(\ref{e2}). Note also that we assume $\nu_s$ to form all Dark Matter,
independently of $m_s$; this can be arranged by an appropriate choice
of $\Delta L$. As a result, our limit (\ref{e3}) is independent of
$m_s$, since the neutrino density $n_s$ and the observational upper
bound on the photon flux both scale like $1/m_s$.}
\be \label{e3}
\tau(\nu_s) > 10^{22} \, {\rm sec}.
\ee
This implies
\be \label{e4}
\sin^2 \theta < 2.9 \cdot 10^{-3} \left( \frac {1 \, {\rm keV}} {m_s}
\right)^5,
\ee
where we have used eq.~(\ref{e1}). This condition imposes a nontrivial
new constraint for $m_s \gsim 2$ keV. Note that this constraint does
not depend on the thermal history of the Universe prior of the
conversion of the lepton asymmetry into sterile neutrinos. It is
depicted by the solid line in Fig.~1.

The constraint (\ref{e3}) has been derived \cite{4} under the
conservative assumption that Dark Matter is distributed uniformly
throughout the Universe. A simple estimate shows that the contribution
from the dark halo of our galaxy alone yields a comparable bound. The
best strategy would probably be to search for the emission of
monoenergetic photons from regions that are known to be rich in Dark
Matter, but have few background sources in the relevant frequency
band. One example might be (the centers of) dwarf galaxies.

Of course, these searches can only be expected to be successful if
$\nu_s$ does indeed form (most of) the Dark Matter in the Universe. We
will now show that the requirement of thermal nonequilibration, the
second condition mentioned above, imposes a severe constraint on the
thermal history of the Universe if Dark Matter is indeed
``cool''. This constraint follows from the requirement that the
present density of {\em thermal} relic $\nu_s$s, which contribute to
hot Dark Matter, should be sufficiently small. Since $\nu_s$ was
relativistic when it decoupled its contribution to the present mass
density of the Universe can be obtained by simple scaling from the
contribution of massive SM neutrinos \cite{5,4}:
\be \label{e5}
\Omega_s^{\rm thermal} h^2 \simeq \frac {m_s} {100 \ {\rm eV}} \cdot
\frac {10.75} {g_*(T_F)},
\ee
where $h$ is today's Hubble constant in units of
100{\thinspace}km$/$(sec$\cdot$Mpc),
and $g_*$ is the number of relativistic degrees
of freedom at the temperature $T_F$ where $\nu_s$ decoupled from the
plasma of SM particles. This temperature is defined by the condition
that the rate of reactions that change the number density $n_s$ of
$\nu_s$ should be equal to the Hubble expansion rate $H$ at that
temperature:
\be \label{e6}
n_s(T_F) \langle v \sigma(\nu_s f \rightarrow \nu f) \rangle(T_F) =
H(T_F).
\ee
Here $f$ stands for any SM fermion or antifermion and $\nu$ for an
active neutrino. The symbol $\langle \cdots \rangle$ denotes the thermal
average. 

\begin{figure}[htb]
\vspace*{-20mm}
\hspace*{5mm}
\mbox{
\epsfig{file=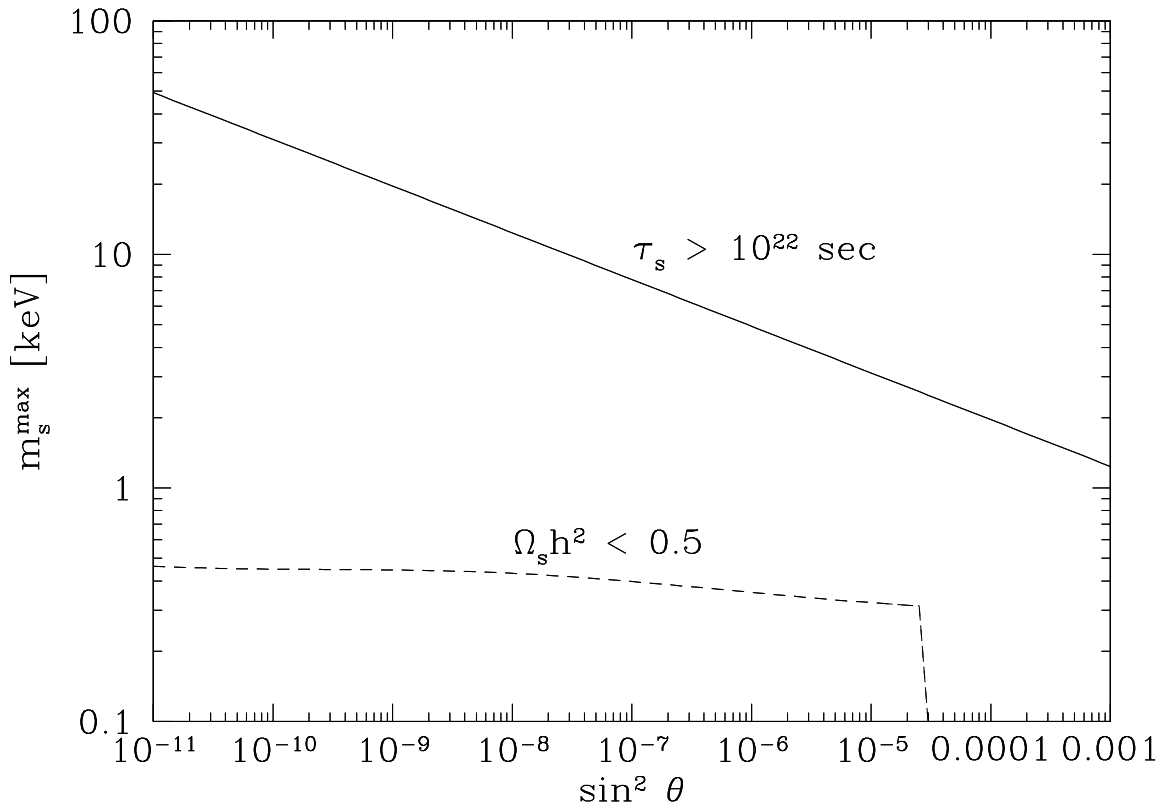,width=0.7\textwidth} }
\vspace*{-20mm}
\caption{\small Upper bounds on the mass $m_s$ of the sterile neutrino
$\nu_s$ as a function of $\sin^2 \theta$, $\theta$ being the effective
mixing angle to SM neutrinos. The solid line comes from the upper
bound on radiative $\nu_s$ decays and is independent of the thermal
history of the Universe. The dashed line follows from the requirement
that {\em thermal} $\nu_s$ relics should not ``overclose'' the
Universe; it has been derived under the assumption that $\nu_s$ was in
thermal equilibrium after inflation. }

\end{figure}

During the radiation dominated era, the Hubble expansion rate was
\cite{4}
\be \label{e7}
H(T) = \frac {\pi T^2} {M_{Pl}} \sqrt{ \frac {g_*}{90} },
\ee
where $M_{Pl} = 2.4 \cdot 10^{18}$ GeV is the reduced Planck
mass. Since by assumption $\nu_s$ was in equilibrium for $T \geq T_F$,
$n_s$ in eq.~(\ref{e6}) can be replaced by the equilibrium density
\cite{4}
\be \label{e8}
n_s = n_s^{\rm eq} = \frac {3 \xi(3)} {2 \pi^2} T^3 = 0.183 T^3.
\ee
Here we have conservatively assumed that $\nu_s$ effectively has only
two degrees of freedom (left-handed $\nu_s$ and right-handed
$\bar{\nu}_s$); if $\nu_s$ is a Dirac particle, this amounts to the
assumption that interactions which change the chirality of $\nu_s$ are
not in equilibrium at $T \simeq T_F$.

We estimate the cross section appearing in eq.~(\ref{e6}) from the
$\nu_\mu e^- \rightarrow \nu_\mu e^-$ scattering cross section. This
is again conservative, since it ignores charged current contributions
to the cross section; moreover, electrons have the smallest (vector)
couplings to the $Z$ among all SM fermions. This gives
\be \label{e9}
\langle v \sigma(\nu_s f \rightarrow \nu f) \rangle = \sin^2 \theta
\cdot \frac {G_F^2}{12 \pi} \left( 3 - 12 \sin^2 \theta_W + 16 \sin^4
\theta_W \right) \cdot \langle s \rangle,
\ee
where $\theta_W$ is the weak mixing angle. We have ignored the
momentum dependence of the $Z$ propagator, since $T_F^2 \ll M_Z^2$ for
all cases of interest. Moreover, when computing the thermal average, we
ignore the Fermi blocking in the final state. The thermal average of
the squared center-of-mass energy $s$ is then given by
\be \label{e10}
\langle s \rangle = 2 \langle E_1 E_2 \rangle = 2 T^2
\left[ \frac{ \int_0^\infty dx x^3/(1 + {\rm e}^x) }
            { \int_0^\infty dx x^2/(1 + {\rm e}^x) } \right]^2 = 19.8
T^2,
\ee
where $E_1$ and $E_2$ are the energies of the two particles in the
initial state in the co-moving frame. Putting everything together, we
have
\be \label{e11}
T_F \simeq 2.2 \, {\rm MeV} \cdot \left( \sin^2 \theta \right)^{-1/3}
\cdot \left[ g_*(T_F) \right]^{1/6}.
\ee
The factor $\left[ g_*(T_F) \right]^{1/6}$ varies between 1.5 for $T_F
\sim 10${\thinspace}MeV and 2.1 for $T_F \sim 10${\thinspace}GeV.
Note that $T_F \lsim 10${\thinspace}GeV for $\sin^2 \theta \gsim 10^{-10}$.

The dashed line in Fig.~1 shows the upper bound on $m_s$ that follows
from eqs.(\ref{e5}) and (\ref{e11}) by requiring that $\nu_s$ should
not ``overclose'' the Universe, $\Omega_s^{\rm thermal} h^2 <
0.5$. The drop at $\sin^2 \theta = 3 \cdot 10^{-5}$ occurs since for
larger mixing angles $T_F$ falls below the temperature of the QCD
phase transition, which we conservatively assumed to be 150{\thinspace}MeV;
this leads to a significant decrease of $g_*(T_F)$ in eq. (\ref{e5}). Note
that the upper bound on $m_s$ should be lowered by at least another
factor of five if $\nu_s$ is to form mostly {\em cool} Dark Matter,
i.{\thinspace}e. if thermal (hot) relics are to form only a minor fraction
of today's $\nu_s$ relic density. The {\em upper} bound on $m_s$ that
follows from the upper bound on the density of thermal $\nu_s$ relics
would then fall slightly {\em below} the {\em lower} bound
on $m_s$ of $\sim$0.1{\thinspace}keV derived in ref.~\cite{1}. 

Simply put, ``cool'' Dark Matter as discussed in refs.~\cite{1,2}
cannot exist if $\nu_s$ ever was in thermal equilibrium. The ``cool''
Dark Matter model is therefore only viable if the reheating
temperature $T_R$ after inflation is (well) below the temperature
(\ref{e11}). For example, certain models of $D$-term inflation have
$T_R \sim 1${\thinspace}GeV \cite{6}; this would leave some parameter
space with $\sin^2 \theta < 10^{-7}$. Note, however, that the density
of thermally produced $\nu_s$ can be significant even if $\nu_s$ never
was in thermal equilibrium. This is analogous to the case of
gravitinos with mass $\gsim 1${\thinspace}keV in supergravity
models. For a given $T_R < T_F$ and a given $\sin^2 \theta$, a
nontrivial upper bound on $m_s$ may therefore still result. Of course,
the requirement of a very low reheating temperature may also make it
difficult to explain the large lepton asymmetry that is required by
the model.

In summary, we have shown that the requirement that the sterile
neutrino be sufficiently long-lived imposes a significant constraint
on the allowed combinations of mass $m_s$ and mixing angle $\theta$
with ordinary (active) neutrinos. Loop induced radiative $\nu_s
\rightarrow \nu \gamma$ decays play a crucial role here. This
constraint is independent of the thermal history of the
Universe. Furthermore, the $\nu_s$ can only form {\em cool} Dark
Matter if, after the end of inflation, it never was in thermal
equilibrium. This imposes a very stringent constraint on the reheating
temperature; e.{\thinspace}g., $T_R < 10${\thinspace}GeV
(1{\thinspace}GeV) for $\sin^2 \theta = 10^{-10} \ (10^{-7})$. These
constraints, together with the requirement of a lepton asymmetry that
is several orders of magnitude larger than the observed baryon
asymmetry, make this model rather unattractive.

\bigskip

\noindent {\bf Acknowledgements:} \\ 
We thank Subir Sarkar for pointing out the importance of radiative
$\nu_s$ decays. This work was supported in part by the
``Sonderforschungsbereich 375--95 f\"ur Astro--Teilchenphysik'' der
Deutschen Forschungsgemeinschaft.

\end{document}